\documentclass[a4paper]{jpconf}
\usepackage{graphicx}
\usepackage{lineno,hyperref}
\usepackage{enumitem}
\usepackage{url}
\usepackage{glossaries}
\usepackage{subfig}
\usepackage{color}
\usepackage{todonotes}
\usepackage{ulem}
\usepackage{amssymb}
\graphicspath{{figures/}}

\begin{document}
\title{The relevance of fluorescence radiation in Cherenkov telescopes}
\author{F. Arqueros, J. Rosado, D. Morcuende, J. L. Contreras}
\address{High Energy Physics Group GAE and UPARCOS. Facultad de Ciencias F\'{i}sicas. Universidad Complutense de Madrid. E-28040 Madrid, Spain.}
\ead{arqueros@ucm.es}

% For the "sloppy" mode: http://tex.stackexchange.com/questions/89354/forcing-math-mode-to-be-on-the-same-line
% \binoppenalty=10000
% \relpenalty=10000

% \newcommand{\tmtexttt}[1]{{\ttfamily{#1}}}
% \newcommand{\apj}{Astrophysical Journal}
% \newcommand{\aap}{A\&A}
% \newcommand{\pasp}{Publications of the ASP}

% \definecolor{navyblue}{rgb}{0.0, 0.0, 0.5}
% \newcommand{\thadd}[1]{\textcolor{navyblue}{[TH: #1]}}

% \begin{frontmatter}
%%%%%%%%%%%%%%%%%%%%%%%%%%%%%%%%%%%%%%%%%%
\begin{abstract}
Cherenkov telescopes are also sensitive to the atmospheric fluorescence produced by the extensive air showers. However this contribution is neglected by the reconstruction algorithms of  imaging air Cherenkov telescopes IACTs and wide-angle Cherenkov detectors WACDs. In this paper we evaluate the fluorescence contamination in the Cherenkov signals from MC simulations in both kinds of Cherenkov telescopes and for some typical observational situations. Results for an observation level of 2200~m a.s.l. are shown. In addition, the feasibility and capabilities of IACTs working as fluorescence telescopes are discussed with the assistance of some geometrical calculations.
\end{abstract}
%%%%%%%%%%%%%%%%%%%%%%%%%%%%%%%%%%%%%%%%%%
\section{Introduction}
\label{sec:introduction}
An extensive air shower (EAS) generates radiation in the optical range due to the Cherenkov effect as well as to the fluorescence emission after the de-excitation of air molecules excited by the charged particles of the shower. Cherenkov and fluorescence photons emitted from any point of an EAS are indistinguishable for a ground Cherenkov telescope because they arrive simultaneously and within the same wavelength range of around $300-450$~nm. On the other hand, fluorescence emission is less efficient than Cherenkov production (e.g., a 1~GeV electron near ground generates in one meter about 30 Cherenkov photons but only about 4 fluorescence photons). In addition, the fluorescence light is emitted isotropically, while Cherenkov radiation is peaked at a small angle of $\sim 1^{\circ}$ around the shower axis. Therefore, the signals recorded by a telescope pointing to the direction of the shower axis (on-axis observation) are dominated by the Cherenkov component as long as the telescope is not far away from the impact point (i.e., within the so-called Cherenkov light pool). This is the case of imaging air Cherenkov telescopes (IACTs) \cite{TheCTAConsortium2013a} used for VHE $\gamma$-ray astronomy. Arrays of wide-angle Cherenkov detectors (WACDs) can also be used as non-imaging telescopes for both $\gamma$- and cosmic-rays \cite{Tluczykont2014}. In this case, the reconstruction of the shower direction relies on the time structure of the front of Cherenkov photons reaching the stations of the array. The fluorescence light also entails a contamination to the recorded signals.
\par
In this work, we have evaluated the fluorescence contamination in Cherenkov telescopes, both IACTs and WACDs, using CORSIKA\cite{corsika}. This code includes the generation and transportation to ground of the Cherenkov photons generated by the showers. Fluorescence emission is not yet included in the official version of CORSIKA, but we have developed an implementation described in \cite{astropart_phys_fluorescence_2018}, which follows a procedure similar to that of the Cherenkov production. With the assistance of this tool, the photon density (photons/m$^2$) of both Cherenkov and fluorescence light at ground has been calculated as a function of the distance to the impact point (intersection of the shower axis with the ground plane) for a sample of EASs initiated by $\gamma$-rays in the energy range 100~GeV -- 1~PeV. Results on fluorescence contamination will be shown in section \ref{sec:results}.
\par
Detection of atmospheric fluorescence is the basis of a well established technique for the study of UHE cosmic rays \cite{Auger2010}. In this case, fluorescence telescopes register air-showers transversely to the shower axis. The working energy range of these telescopes is limited by several factors like the night sky background and the telescope field of view. Although IACTs are designed for the detection of Cherenkov light from $\gamma$-ray sources, in principle, they could also be used with minor modifications in ``fluorescence mode" for the detection of $\gamma$- or cosmic-ray showers \cite{ICRC2015}. In section \ref{sec:IACT_fluorescence} we will discuss on geometrical aspects of interest for this topic. 
%
%%%%%%%%%%%%%%%%%%%%%%%%%%%%%%%%%%%%%%%%%%
\section{MC simulations}
\label{sec:MC_simul}
The main options and parameters used in CORSIKA for these simulations have been the following. For hadronic interactions the QGSJET-01C and GHEISHA 2002d models have been chosen for high- and low-energy interactions, respectively. The energy thresholds were set to 0.3~GeV for hadrons, 0.1~GeV for muons and 0.02~GeV for the electromagnetic component. The observation level was set to an altitude of 2200~m a.s.l. and photons reaching a square $5 \times 5$~km$^2$ centered at the impact point were stored.
\par
\begin{figure}[h]
\includegraphics[width=17.5pc]{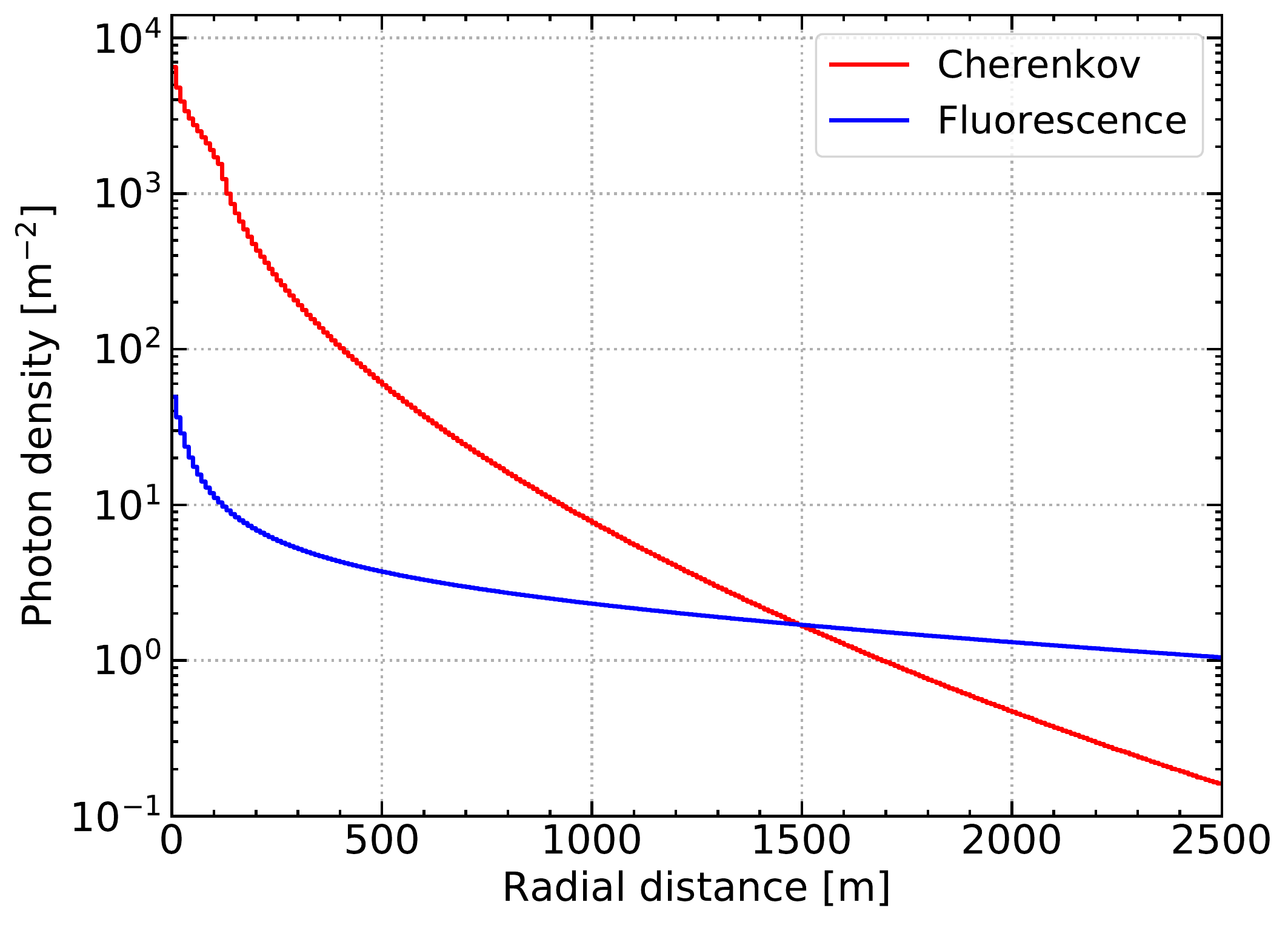}\hspace{2pc}%
\begin{minipage}[b]{17.5pc}\caption{\label{fig:10TeV_vertical_radial}Average
radial distributions of Cherenkov (red) and fluorescence (blue) light for vertical 10 TeV $\gamma$-ray showers. All photons (without any angular cut)
have been used here.}
\end{minipage}
\end{figure}

 
MC data were analyzed with dedicated Python scripts. The detection area was discretized into a dense grid and the number of photons in each grid element was counted. As an example, the average radial distribution of both Cherenkov and fluorescence photon densities for vertical 10~TeV $\gamma$-ray showers is shown in figure \ref{fig:10TeV_vertical_radial}. It can be observed that Cherenkov light concentrates on the pool region, which extends up to a radial distance of $\sim 120$~m in this case, while the photon density sharply drops outside. On the other hand, the fluorescence photon density decreases much less steeply with increasing core distance due to the isotropic nature of the fluorescence emission. In fact, the fluorescence photon density becomes larger than the Cherenkov one at very large core distance.
\par
%
%%%%%%%%%%%%%%%%%%%%%%%%%%%%%%%%%%%%%%%%%%
\section{Fluorescence contamination in Cherenkov telescopes}
\label{sec:results}
\begin{figure}[h]
\begin{minipage}{17.5pc}
\includegraphics[width=17.5pc]{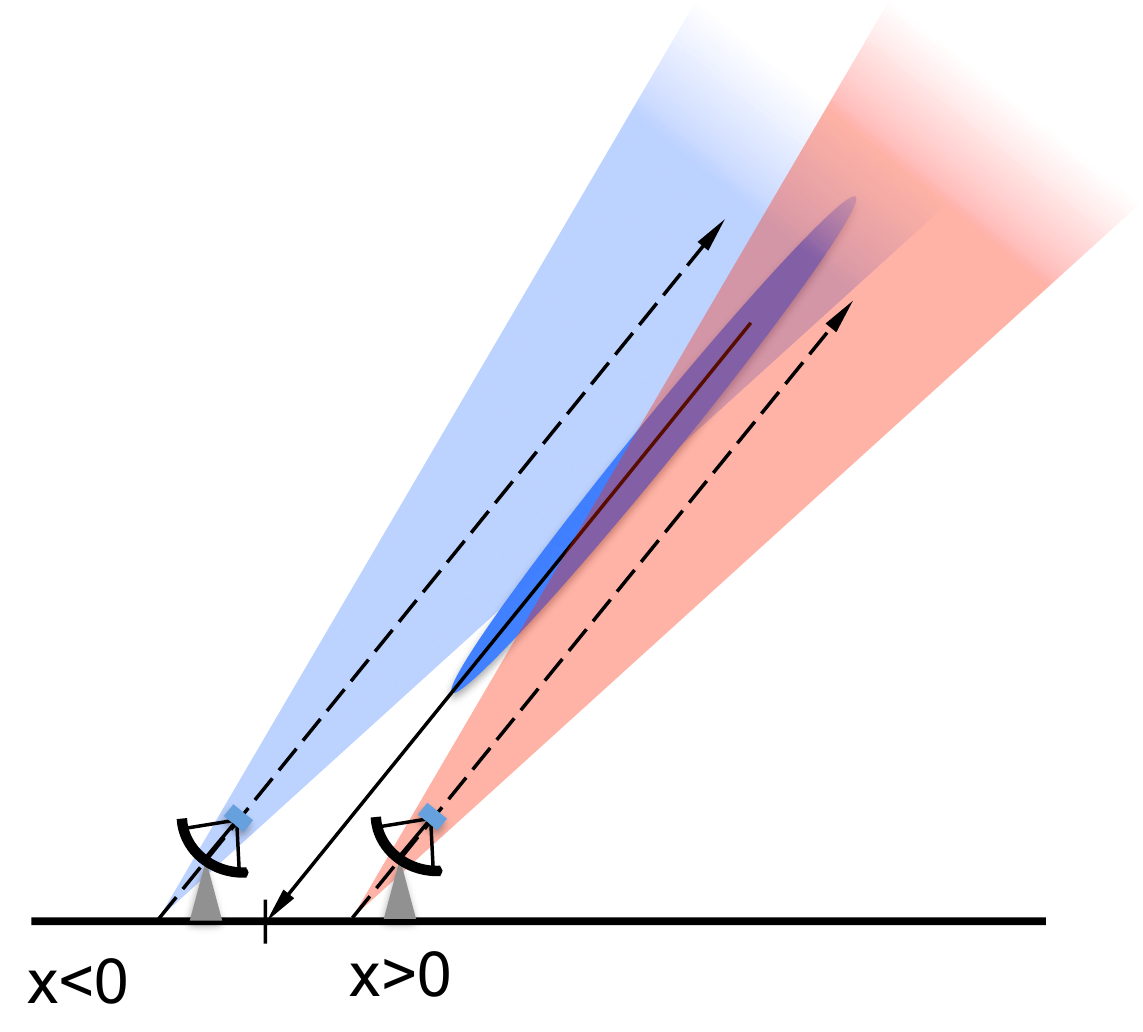}
\caption{\label{fig:geometry_IACT}Schematic representation of the geometry for IACT observations.}
\end{minipage}\hspace{2pc}%
\begin{minipage}{17.5pc}
\includegraphics[width=17.5pc]{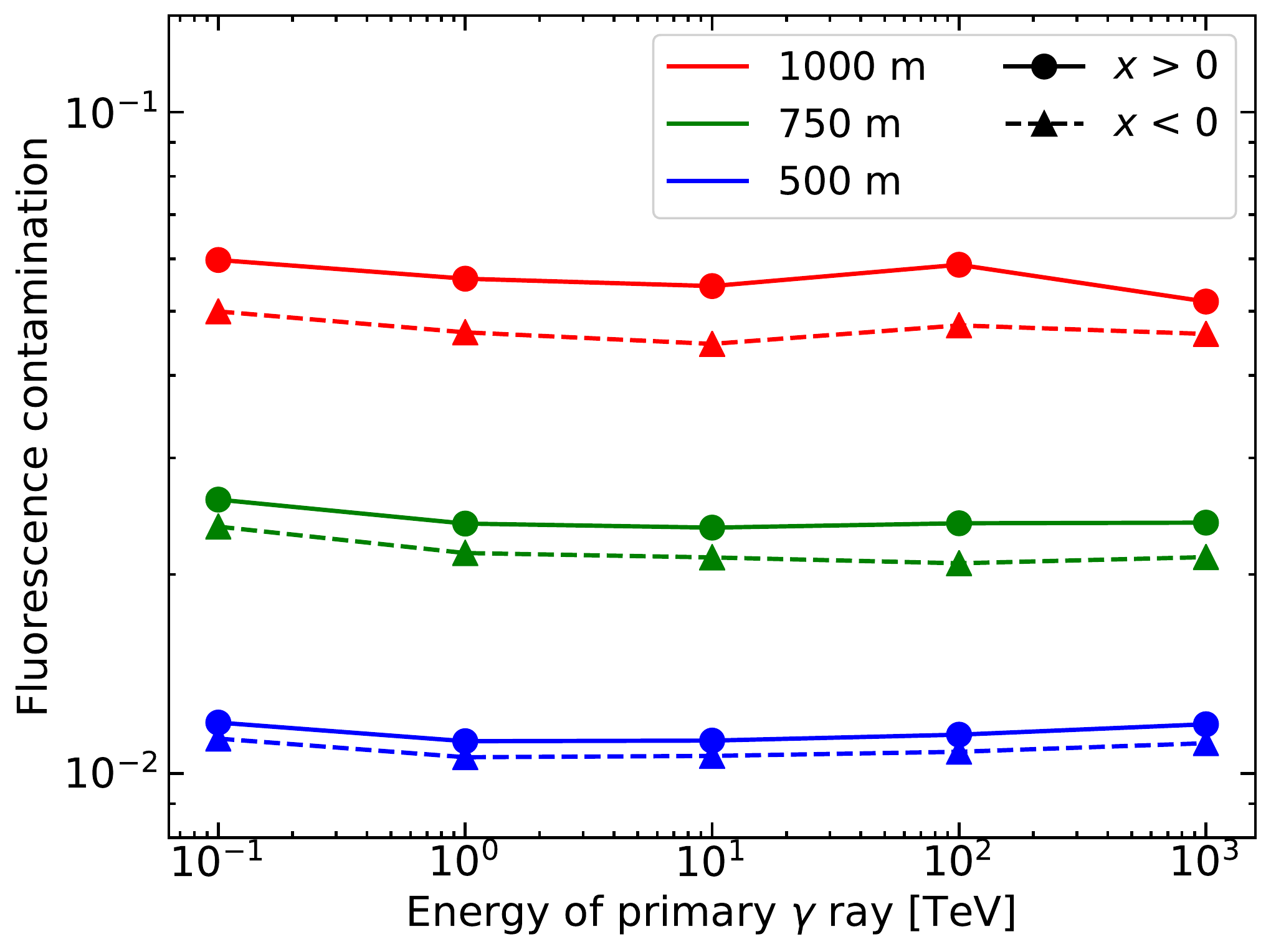}
\caption{\label{fig:Energy_IACT}$R_{\rm FC}$ versus $E$ for $\gamma$-ray showers of $20^\circ$ zenith angle at three distances of the impact point.}
\end{minipage} 
\end{figure}

\begin{figure}[h]
\begin{minipage}{17.5pc}
\includegraphics[width=17.5pc]{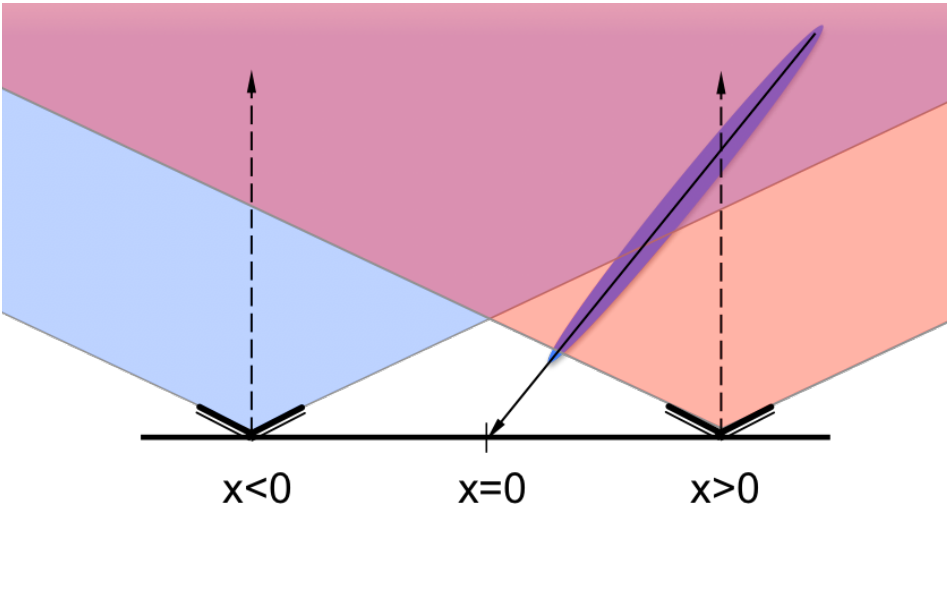}
\caption{\label{fig:geometry_WACD}Schematic representation of the geometry for WACDs.}
\end{minipage}\hspace{2pc}%
\begin{minipage}{17.5pc}
\includegraphics[width=17.5pc]{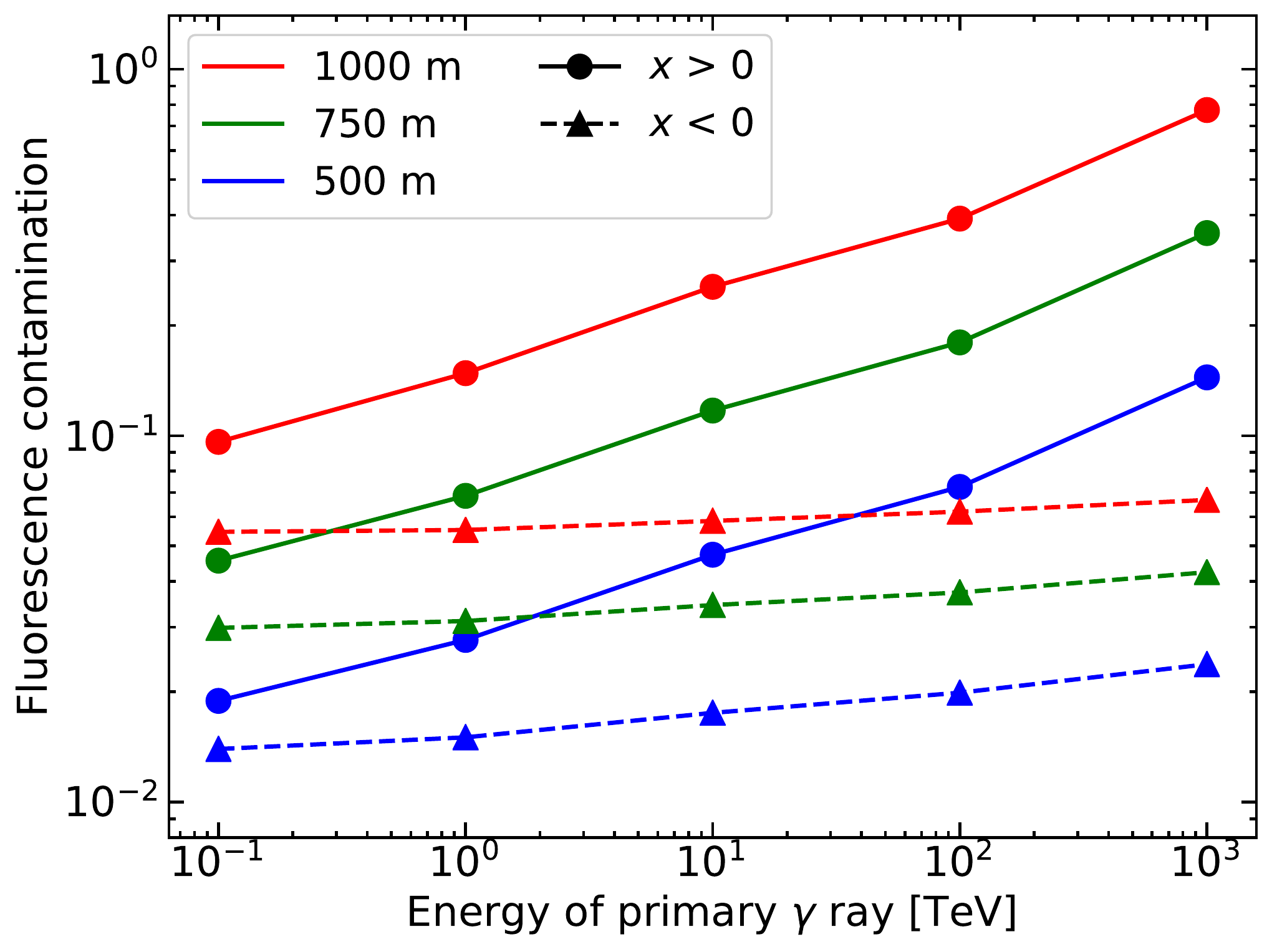}
\caption{\label{fig:Energy_WACD}The same as figure \ref{fig:Energy_IACT} for WACDs.}
\end{minipage} 
\end{figure}

Two different observational techniques have been studied: IACTs working in on-axis mode with a FoV of $10^\circ$ in diameter (see figure \ref{fig:geometry_IACT}) and WACDs pointing to the zenith for which we have assumed a FoV of $60^\circ$ in diameter (see figure \ref{fig:geometry_WACD}). The pointing direction and FoV of these telescopes have been simulated by applying the appropriate cuts on the arrival angle of the photons. No cut in the arrival time was applied for this work.
\par 
For inclined showers we have calculated the photon densities only along the $x$ axis, defined as the projection of the shower axis on the ground plane. The fluorescence contamination $R_{\rm FC}$ at a given position on ground is defined in this work as the ratio between the fluorescence and Cherenkov photon densities and it has been evaluated for both IACTs and WACDs as a function of the $x$ distance, the shower energy $E$ and zenith angle $\theta$. 
\subsection*{IACT case}
\label{sec:IACT}
The fluorescence contribution is strongly suppressed in telescopes with a narrow FoV in such a way that the contamination is negligible within the light pool. Results for $20^\circ$ $\gamma$-showers are displayed in figure \ref{fig:Energy_IACT} showing that $R_{\rm FC}$ has a very weak dependence with energy. It is also nearly independent of the $x$ sign as expected from symmetry reasons (see figure \ref{fig:geometry_IACT}).
\par
The $\theta$ dependence of $R_{\rm FC}$ has been found to be negligible if evaluated in a plane perpendicular to the shower axis. However, for a horizontal plane (i.e., the ground), the contamination for a fixed $x$ value decreases with the shower inclination, becoming negligible beyond about $40^\circ$.
\par
It should be stressed that the fluorescence contamination becomes increasingly significant at large core distances (e.g., $>5$\% at 1000~m), as can be observed in figure \ref{fig:Energy_IACT}. Thus ignoring this effect would have an impact on VHE $\gamma$-ray observations carried out by arrays of IACTs that collect measurable signals at large core distances (e.g., CTA \cite{TheCTAConsortium2013a}).
\subsection*{WACD case}
\label{sec:WACD}
Figure \ref{fig:Energy_WACD} shows the results for $20^\circ$ $\gamma$-showers. Notice that $R_{\rm FC}$ depends significantly on the $x$ sign, because of the evident asymmetry of this configuration (see figure \ref{fig:geometry_WACD}). For a given $|x|$ value, the detection of the isotropic fluorescence light is favored in the positive side, while this is not the case for the directional Cherenkov radiation. The effect becomes more important as the shower energy increases. Besides, we have found that $R_{\rm FC}$ is weakly dependent of the shower inclination in the $x$ positive side, while decreases with $\theta$ in the negative one.
\par
Our results indicate that the signals of a detector located at large distance from the impact point can be contaminated with a non-negligible fraction of fluorescence light. In particular, the contamination may be as large as 45\% (i.e., $R_{\rm FC} = 0.8$) at $x=1000$~m in the PeV energy range.
\section{IACTs in fluorescence mode}
\label{sec:IACT_fluorescence}
\begin{figure}[h]
\includegraphics[width=18pc]{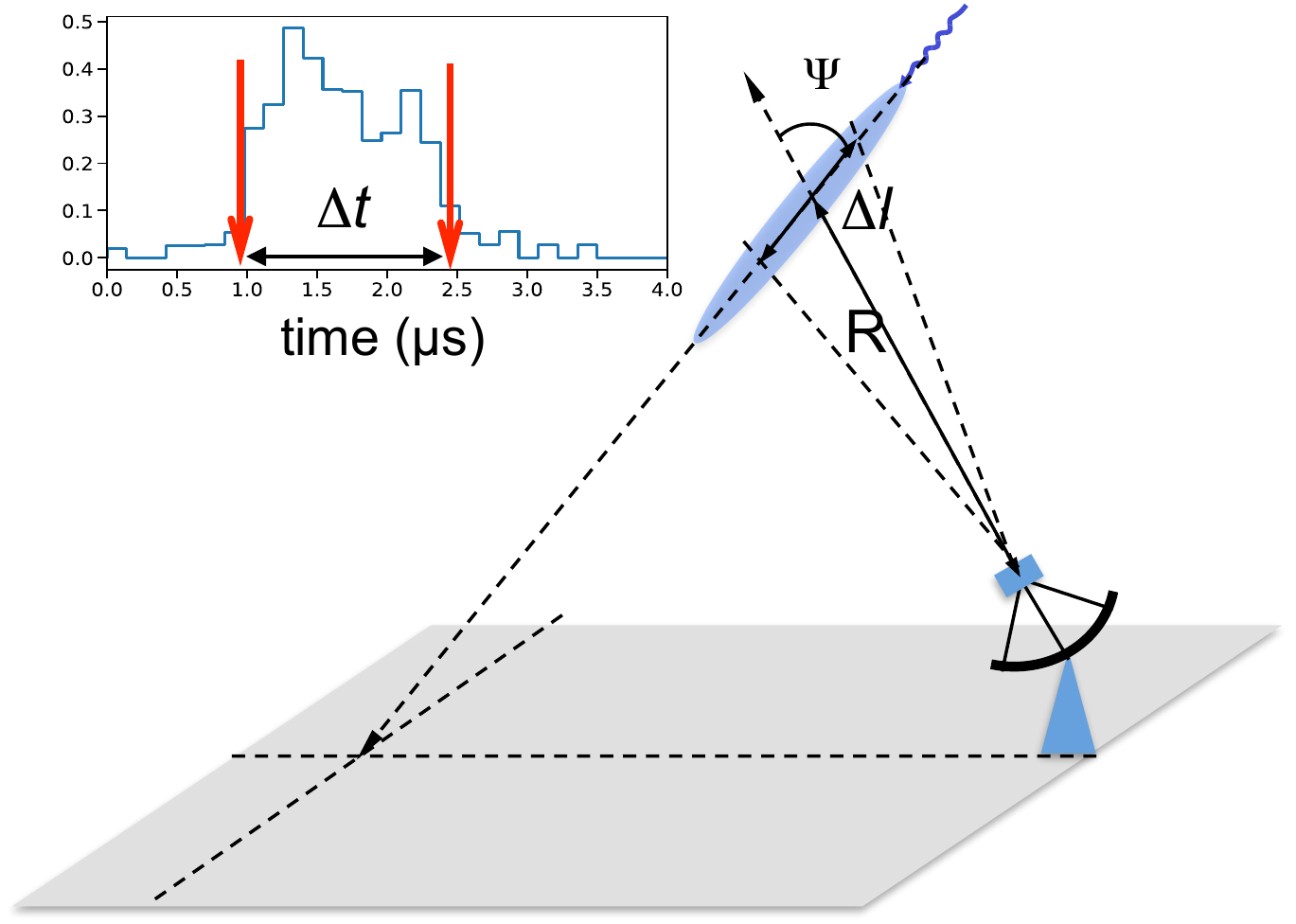}\hspace{2pc}%
\begin{minipage}[b]{18pc}\caption{\label{fig:IACT_fluor_geometry}Geometrical configuration in the fluorescence technique. For simplification we will assume that the optical axis crosses the shower axis and thus the track coincides with a diameter of the camera. $R$ is the shower-to-telescope distance measured along the optical axis, $\Psi$ is the angle between both axes and $\Delta l$ is the length of the track within the telescope FoV. $\Delta t$ is the pulse duration in the camera.}
\end{minipage}
\end{figure}

\begin{figure}[h]
\begin{minipage}{17.5pc}
\includegraphics[width=17.5pc]{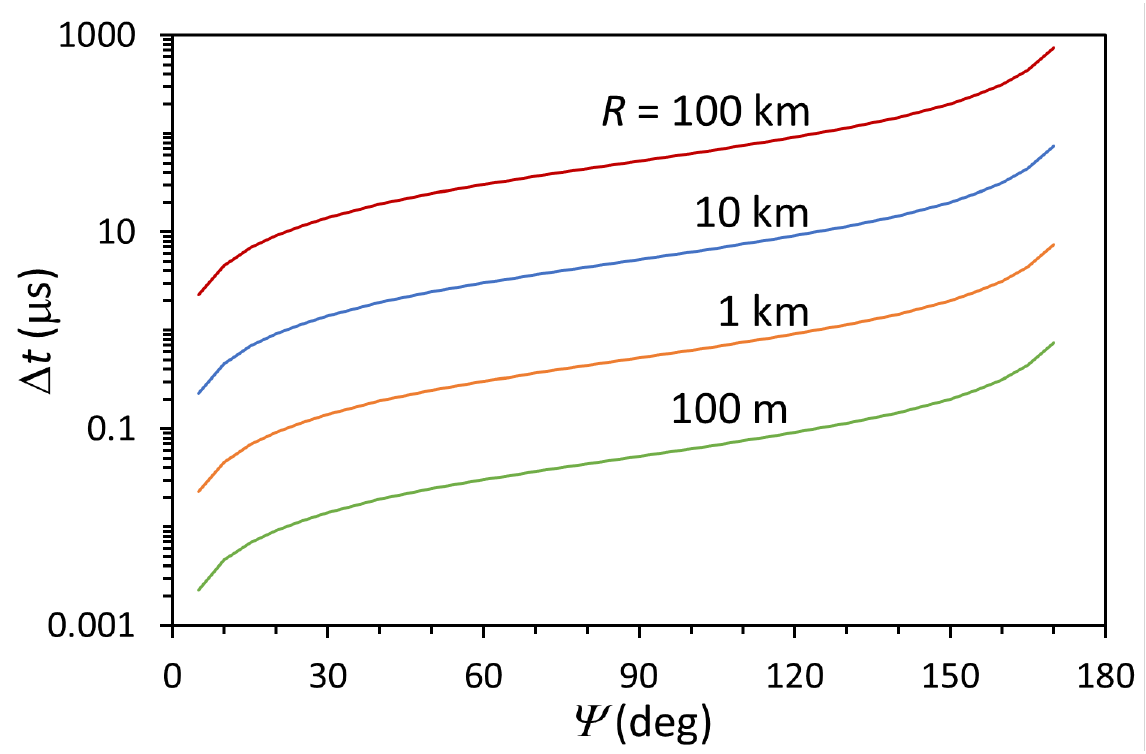}
\caption{\label{fig:IACT_fluor_time}Duration of fluorescence pulse traversing a telescope FoV of $9^\circ$ versus $\Psi$ for several values of $R$.}
\end{minipage}\hspace{2pc}%
\begin{minipage}{17.5pc}
\includegraphics[width=17.5pc]{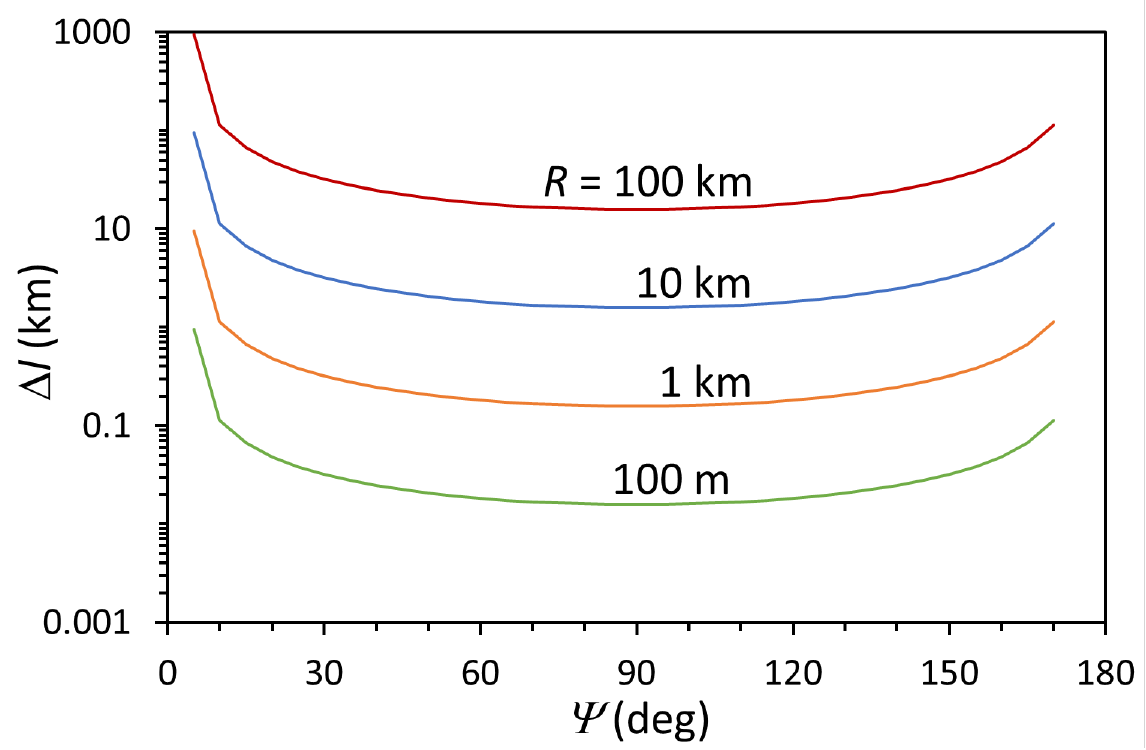}
\caption{\label{fig:IACT_fluor_length}The same as Figure \ref{fig:IACT_fluor_time} for the length of the track observed by the telescope.}
\end{minipage} 
\end{figure}

In a previous paper \cite{ICRC2015}, we discussed for the first time about the feasibility of a large array of IACTs, like CTA 
\cite{TheCTAConsortium2013a}, to work simultaneously as a fluorescence observatory. In this scenario the telescopes should fulfill several requirements related to sensitivity (e.g., the trigger configuration).
Also there are some general geometrical constraints that we discuss next.
Figure \ref{fig:IACT_fluor_geometry} is a sketch showing the definition of the relevant geometric parameters.
In the first place, the effective time window of the camera has to be wide enough to record the passage of the shower track through the telescope FoV (i.e., larger than $\Delta t$). In addition, the length $\Delta l$ of the track observed by the telescopes should be sufficient for reconstruction purposes. The results of simple geometrical calculations for a telescope with a FoV of $9^\circ$ in diameter are displayed in figures \ref{fig:IACT_fluor_time} and \ref{fig:IACT_fluor_length}.
From these results, we infer that it would be possible to record a large fraction ($\Delta l>5$~km) of the track of showers developed at tens of km distances in case that a time window of tens of $\mu$s is available. A time window of about 1~$\mu$s could be sufficient to reach showers at distances of a few km (full track at $\Psi<30^\circ$). Even though IACTs are designed to record Cherenkov flashes that are much shorter ($< 100$~ns), some camera types could be adapted to register traces over time intervals of some $\mu$s \cite{FlashCam}. A modest time resolution would be enough to store those traces, 
keeping the high time resolution for prompt Cherenkov signals.
\par
Stereoscopic observation by several IACTs of the array would be necessary for an accurate shower reconstruction. In addition, this may allow the registration of the full shower track even though each telescope only images a small $\Delta l$ portion of the shower because of its limited FoV.
\par
IACTs would have many advantages in comparison with classical fluorescence telescopes. The small angular size of the pixels of the camera (typically a factor of 10 smaller) leads to a much improved angular resolution. This feature together with the large size and good quality of their mirrors would lower the threshold for detecting light pulses over the night sky background.
\par
The exploitation of fluorescence signals could be applied to several fields. In regard with VHE gamma-ray astronomy, it would be possible to enlarge the effective area of IACTs by the combined detection of Cherenkov and fluorescence light in showers with low $\Psi$ angle. This would require modest extensions of the time windows. If a time window of microseconds is available, an array of IACTs covering a wide FoV could be used in pure fluorescence mode for the stereoscopic detection of EASs. It would allow for registering cosmic-ray showers with unprecedented resolution in the radial distribution along the whole longitudinal development, providing very valuable information for studies of hadronic interactions in atmospheric showers. Large time windows could also open the detection of extremely energetic showers at hundreds of kilometers, yielding huge effective areas.
\par
We plan to make a quantitative evaluation of the feasibility of these potential applications, including the calculation of energy thresholds, effective areas, primary discrimination, etc. These tasks will require detailed simulations for which our implementation of the fluorescence emission in CORSIKA will be the ideal tool.
\section*{Acknowledgements}
We gratefully acknowledge support from Spanish MINECO (contracts FPA2015-69210-C6-3-R and FPA2017-82729-C6-3-R) and the European Commission (E.U. Grant Agreement 653477). D. Morcuende acknowledges a predoctoral grant UCM-Harvard University (CT17/17--CT18/17) from \textit{Universidad Complutense de Madrid}.
%%%%%%%%%%%%%%%%%%%%%%%%%%%%%%%%%%%%%%%%%%

%
\end{document}